\documentclass[10pt]{article}
\usepackage{geometry}                
\usepackage{graphicx}
\usepackage{amssymb}
\usepackage{epstopdf}
\usepackage{color}
\usepackage{subfigure}
\usepackage{feynarts}
 \usepackage{amsmath, amsthm}
\newcommand{\ttt}{t\bar{t}}
\newcommand{\alphas}{\alpha_{s}}
\newcommand{\afb}{A_{FB}}
\newcommand{\alab}{A_{FB}^{p\bar{p}}}
\newcommand{\att}{A_{FB}^{t\bar{t}}}

\begin{document}

\thispagestyle{empty}
\setcounter{page}{0}
\def\thefootnote{\fnsymbol{footnote}}

{\textwidth 15cm

\begin{flushright}
MPP-2011-82\\
\end{flushright}

\vspace{2cm}

\begin{center}

{\Large\sc {\bf The electroweak contribution to the top quark

forward-backward asymmetry at the Tevatron}}

\vspace{2cm}

\sc{W. Hollik\footnote{hollik@mppmu.mpg.de}} \rm  ~and \sc{D. Pagani\footnote{pagani@mppmu.mpg.de}}\rm
\vspace{1cm}

     Max-Planck-Institut f\"ur Physik \\
     (Werner-Heisenberg-Institut)\\
     D-80805 M\"unchen, Germany

\end{center}

\vspace*{2cm}

\begin{abstract}
\noindent
The electroweak contributions to the forward-backward asymmetry 
in the production of top-quark pairs at the Tevatron are 
evaluated at $\mathcal{O}(\alpha^2)$ and 
$\mathcal{O}(\alpha\alphas^{2})$. We perform a detailed analysis 
of all partonic channels that produce an asymmetry 
and combine them with the QCD contributions. 
They
provide a non-negligible fraction
of the QCD-induced asymmetry with the same overall sign, thus
enlarging the Standard Model prediction
and diminishing the observed deviation.
For the observed mass-dependent forward-backward asymmetry 
a $3 \sigma$ deviation still remains at an invariant-mass cut
of $M_{\ttt} > 450$~GeV.   
\end{abstract}

}
\def\thefootnote{\arabic{footnote}}
\setcounter{footnote}{0}
\newpage

\section{Introduction}

The investigation of the top quark at the Tevatron has substantially contributed to
precision tests of QCD and the electroweak theory.
Besides the valuable set of top-quark observables like mass, width, cross section,
which are fully consistent with the SM,  the measured forward-backward asymmetry $A_{FB}$
of top-pair production~\cite{Aaltonen:2008hc,Park:2008rw} is larger than 
expected from  the Standard Model (SM) prediction 
and has led to speculations on the presence of possible new physics.

Two options for the forward-backward asymmetry have been used in the experimental analysis,  
with the definitions 
\begin{equation}
\label{Afbtt}
  A_{FB}^{t\bar{t}}=
  \frac{\sigma(\Delta y > 0) - \sigma(\Delta y < 0)}{\sigma(\Delta y > 0) + \sigma(\Delta y < 0)}
\end{equation}
and
\begin{equation}
\label{Afblab}
 \alab=\frac{\sigma(y_{t} > 0) - \sigma(y_{t}< 0)}{\sigma(y_{t}> 0) + \sigma(y_{t}< 0)}
\end{equation}
given in~\cite{Aaltonen:2011kc} reporting the recent CDF result.
$\Delta y$ is defined as the difference between the rapidity $y_{t}$ and $y_{\bar{t}}$ of $t$ 
and $\bar{t}$ where the proton direction defines the beam axis.
$\Delta y$ (not  $y_{t}$) is invariant under a boost along the beam axis, 
thus it is the same in the partonic and in the hadronic rest frame.

The recent values for the inclusive asymmetry obtained by CDF~\cite{Aaltonen:2011kc} are
\begin{eqnarray}\label{CDF}
A_{FB}^{t\bar{t}}=0.158\pm0.075, \\
\nonumber 
 \alab=0.150\pm0.055 .
\end{eqnarray}
The LO predictions of the Standard Model originate from 
NLO QCD contributions to the differential cross section for
$\ttt$ production that are antisymmetric under 
charge conjugation~\cite{Kuhn:1998jr,Kuhn:1998kw}, yielding values  
for $\att(\alab)$ around $7\%(5\%)$ (see e.g.~\cite{Bernreuther:2010ny}). 
The observed difference between the measurement and the prediction has inspired
quite a number of theoretical papers proposing various new physics mechanisms as potential additional
sources for the forward-backward asymmetry 
(see for example~\cite{Rodrigo:2010gm,Shu:2011au} and references therein).

The importance of identifying signals from possible new physics requires a thorough discussion 
of the SM prediction and the corresponding uncertainty. At present, the theoretical accuracy
is limited by the missing NNLO contribution from QCD to the antisymmetric part of the
$\ttt$ production cross section. Besides the strong interaction, the electroweak interaction
gives rise to further contributions
to the $\ttt$ forward-backward asymmetry, 
through photon and $Z$ exchange at the tree level as well as through interference between
QCD and electroweak amplitudes at NLO in both interactions. Although smaller in size,
they are not negligible, and 
a careful investigation is an essential ingredient for an improved theoretical prediction.

In this paper we perform a detailed analysis of the electroweak contributions  
to the forward-backward asymmetry in $\ttt$ production based on the evaluation 
of all partonic channels that produce an asymmetry both at the tree level and at NLO,
and combine them with the QCD contributions.
We apply the calculation to both types of asymmetries given above in \eqref{Afbtt}  
and \eqref{Afblab}. Moreover, we present results for $\att$ also with a cut
$M_{\ttt}>450 \text{ GeV}$ on the invariant $\ttt$ mass as well as with a rapidity cut
$|\Delta y|>1$, for comparison with the experimental values given in~\cite{Aaltonen:2011kc},   
\begin{equation}\label{minvy}
A_{FB}^{t\bar{t}}(M_{\ttt}\ge450 \text{ GeV})=0.475\pm0.114, \qquad 
A_{FB}^{t\bar{t}}(|\Delta y|\ge1)=0.611\pm0.256 ,
\end{equation}
where in particular the result for the high invariant-mass cut exhibits the largest deviation 
from the QCD prediction.

\section{Calculational basis}

At leading order the production of $\ttt$ pairs in $p\bar{p}$ collisions originates, via the strong interaction,
from the partonic processes  $q\bar{q}\rightarrow\ttt$ and $gg\rightarrow\ttt$, which 
yield the $\mathcal{O}(\alphas^{2})$ of the (integrated) cross section, i.e.\ 
the denominator of $\afb$ in \eqref{Afbtt} and \eqref{Afblab}.
The antisymmetric cross section, the numerator of $\afb$, 
starts at $\mathcal{O}(\alphas^{3})$ and gets
contributions from  $q\bar{q}\rightarrow\ttt(g)$ with $q=u,d$ (the processes from
other quark species, after convolution with the parton distributions and summation,
are symmetric under  $y_{t}\rightarrow- y_{t}$ and thus do
not contribute to $\afb$) as well as from  
$qg\rightarrow\ttt q$ and $\bar{q}g\rightarrow\ttt \bar{q}$.

Writing the numerator and the denominator of $\afb$ 
(for either of the definitions  (\ref{Afbtt}) and (\ref{Afblab}))
in powers of $\alphas$ we obtain
\begin{equation}\label{powers}
\afb=\frac{N}{D}=\frac{\alphas^{3} N_{1}+\alphas^{4} N_{2}+\cdots}{\alphas^{2} D_{0}+\alphas^{3} D_{1}+\cdots}=\frac{\alphas}{D_{0}}(N_{1}+\alphas(N_{2}-N_{1}D_{1}/D_{0}))+\cdots.
\end{equation}
The terms up to one-loop ($D_{0},D_{1},N_{1}$) have been calculated \cite{Gluck:1977zm,Combridge:1978kx,Babcock:1977fi,Hagiwara:1978hw,Jones:1977di,Georgi:1978kx}, \cite{Nason:1987xz,Altarelli:1988qr,Beenakker:1988bq,Melnikov:2009dn}, \cite{Kuhn:1998kw},
whereas only some parts of $N_{2}$ are currently known \cite{Ahrens:2011mw,Ahrens:2011uf}.  
The inclusion of the $N_{1}D_{1}/D_{0}$ term without $N_{2}$ would hence be 
incomplete, and we have chosen to use only
the lowest order cross section in the denominator and the $\mathcal{O}(\alphas^{3})$
term in the numerator, as done in \cite{Kuhn:1998kw}.

Rewriting $N$ and $D$ to include the EW contributions yields the following 
expression for the leading terms,
\begin{equation}\label{powersew}
\afb=\frac{N}{D}=\frac{\alpha^{2} \tilde{N}_{0}+\alphas^{3} N_{1}+\alphas^{2}\alpha \tilde{N}_{1}+\alphas^{4} N_{2}+\cdots}{\alpha^{2} \tilde{D}_{0}+\alphas^{2} D_{0}+\alphas^{3} D_{1}+\alphas^{2}\alpha \tilde{D}_{1}+\cdots}=\alphas\frac{N_{1}}{D_{0}}+\alpha\frac{\tilde{N}_{1}}{D_{0}}+\frac{\alpha^{2}}{\alphas^{2}}\frac{\tilde{N}_{0}}{D_{0}}+\cdots
\end{equation}
where the incomplete $\mathcal{O}(\alphas^{2})$ part has been dropped.
In the following we (re-)evaluate the three contributions on the r.h.s.\ 
of~\eqref{powersew}.
\begin{figure}[!h]
\centering
\input{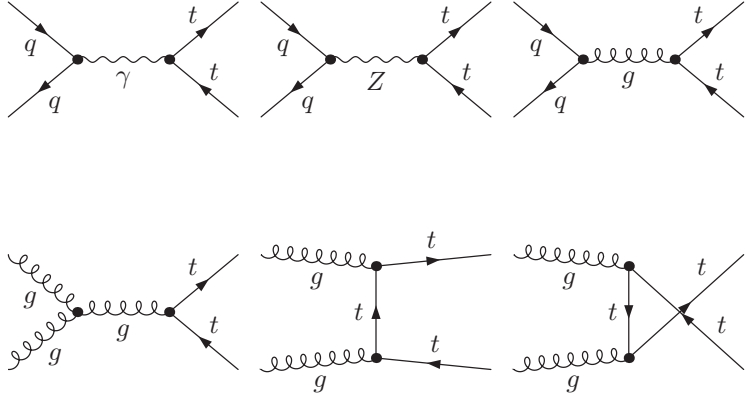}
\caption{Electroweak and QCD Born diagrams}
\label{bornd}
\end{figure}

Figure~\ref{bornd} contains all the tree level diagrams for the partonic  subprocesses 
$q\bar{q}\rightarrow\ttt$ and $gg\rightarrow\ttt$ 
(Higgs exchange is completely negligible).
The squared terms $|\mathcal{M}_{q\bar{q}\rightarrow g \rightarrow\ttt}|^{2}$ 
and $|\mathcal{M}_{g\bar{g}\rightarrow \ttt}|^{2}$  yield $D_{0}$ of the LO cross section;
the  $\mathcal{O}(\alpha^{2})$ terms arise from
$|\mathcal{M}_{q\bar{q}\rightarrow\gamma\rightarrow\ttt}+\mathcal{M}_{q\bar{q}\rightarrow  Z \rightarrow\ttt}|^{2}$,
which generate a purely-electroweak antisymmetric differential cross section,
in the parton cms given by
 \begin{eqnarray}\label{gamma-z}
\frac{d\sigma_{asym}}{d\cos\theta}=2\pi \alpha^{2}\cos\theta\, \Big(1-\frac{4m_{t}^{2}}{s}\Big)\Big[\kappa\frac{Q_{q}Q_{t}A_{q}A_{t}}{(s-M_{Z}^2)}+2\kappa^{2}A_{q}A_{t}V_{q}V_{t}\frac{s}{(s-M_{Z}^2)^2}\Big], \\
\kappa=\frac{1}{4\sin^{2}(\theta_{W})\cos^{2}(\theta_{W})},\qquad
V_{q}=T^{3}_{q}-2Q_{q}\sin^{2}(\theta_{W}), \qquad A_{q}=T^{3}_{q} . \nonumber
\end{eqnarray}
In $\afb$ \eqref{powersew} this leads to the term $\tilde{N}_0$. The complementary
symmetric cross section provides the $\tilde{D}_0$ term in the denominator,
which does not contribute in the order under consideration.
Interference of $q\bar{q}\rightarrow\gamma,Z\rightarrow\ttt$ and
$q\bar{q}\rightarrow g \rightarrow\ttt$ 
is zero  because of the color structure\footnote{For $q\bar{q}\rightarrow t\bar{t}$ there are also
 $\mathcal{O}(\alpha)$ $W$-mediated $t$-channel diagrams 
 with $q=d,s,b$, but they are
 strongly suppressed by the CKM matrix or by parton distributions ($q=b$).}.

\begin{figure}[!h]
\centering
\input{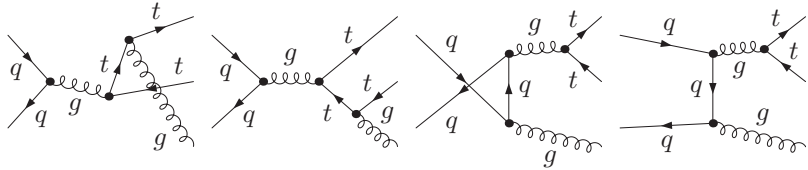}
\caption{Real emission of gluons at $\mathcal{O}(\alphas^{3})$}
\label{realg}
\end{figure}

The $\mathcal{O}(\alphas^{3})$ terms that contributes to $N$ arise from four
classes of partonic processes: 
$q\bar{q}\rightarrow\ttt$, $q\bar{q}\rightarrow\ttt g$, 
$qg\rightarrow\ttt q$ and $\bar{q}g\rightarrow\ttt \bar{q}$.
In the first case the origin is the interference of QCD one-loop and
Born amplitudes; the other processes correspond to real-particle emissions.
\begin{figure}[!h]
\centering
\input{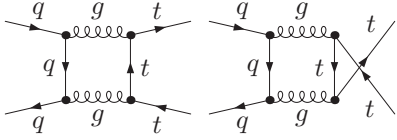}
\caption{QCD box diagrams}
\label{boxqcd}
\end{figure}
All one-loop  vertex corrections and self-energies do not generate any
asymmetric term, hence only the box diagrams (Figure~\ref{boxqcd}) are relevant. 
The box integrals are free of ultraviolet and collinear divergences, but they involve
infrared singularities
which are cancelled after adding the integrated real-gluon emission
contribution $q\bar{q}\rightarrow\ttt g$, shown in Figure~\ref{realg}.
For the corresponding relevant gluon-radiation part only  the interference 
of initial and final state gluon radiation has to  be taken into account,
yielding another antisymmetric cross section. 
The processes of real-quark radiation
$qg\rightarrow\ttt q$ and $\bar{q}g\rightarrow\ttt\bar{q}$ 
yield contributions to $\afb$ which are numerically 
not important~\cite{Kuhn:1998kw}.

\begin{figure}[!h]
\centering
\input{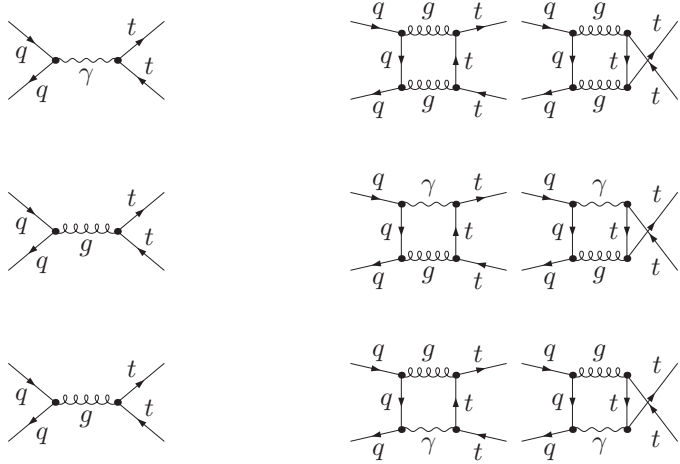}
\caption{Different ways of QED--QCD interference at $\mathcal{O}(\alphas^{2}\alpha)$}  
\label{bornbox}
\end{figure}

In order to analyze the electroweak $\mathcal{O}(\alphas^{2}\alpha)$ terms, 
it is useful to separate the QED contributions involving photons from the weak
contributions with $Z$ bosons.
In the QED sector we obtain the $\mathcal{O}(\alphas^{2}\alpha)$ contributions
to $N$ from three classes of partonic processes:
$q\bar{q}\rightarrow\ttt$, $q\bar{q}\rightarrow\ttt g$ and $q\bar{q}\rightarrow\ttt \gamma$.
The first case is the virtual-photon contribution, 
which can be obtained from the QCD analogue, namely the $\mathcal{O}(\alphas^{3})$ 
interference of box and tree-level amplitudes,
by substituting successively each one of the three internal gluons by a photon,
as displayed in Figure~\ref{bornbox}.

The essential differences between the calculation of the $\mathcal{O}(\alphas^{3})$
and of QED $\mathcal{O}(\alphas^{2}\alpha)$ terms  
are the coupling constants and the appearance of the $SU(3)$ generators in the
strong vertices. 
Summing over color in the final state and averaging in the initial state we find 
for the virtual contributions to the antisymmetric cross section the following ratio,
\begin{equation}\label{rappqcdew}
\frac{\overline{|\mathcal{M}^{\ttt}|}^{2}_{\mathcal{O}(\alphas^{2}\alpha),asym}}{\overline{|\mathcal{M}^{\ttt}|}^{2}_{\mathcal{O}(\alphas^{3}),asym}}=\frac{\overline{2Re\big(\mathcal{M}^{\ttt}_{\mathcal{O}(\alpha)}\mathcal{M}^{\ttt\quad*}_{\mathcal{O}(\alphas^{2})}\big)}_{asym}+2Re\overline{\big(\mathcal{M}^{\ttt}_{\mathcal{O}(\alphas)}\mathcal{M}^{\ttt\quad*}_{\mathcal{O}(\alphas\alpha)}\big)}_{asym}}{\overline{2Re\big(\mathcal{M}^{\ttt}_{\mathcal{O}(\alphas)}\mathcal{M}^{\ttt\quad*}_{\mathcal{O}(\alphas^{2})}\big)}_{asym}}=\frac{F^{\ttt}_{QED}(\alphas,\alpha,Q_{t},Q_{q})}{F^{\ttt}_{QCD}(\alphas)}
\end{equation}
that can be expressed in terms of two factors
$F^{\ttt}_{QED}$ and $F^{\ttt}_{QCD}$ depending only on coupling constants
and color traces,
\begin{subequations}
	\label{factors}
	\begin{eqnarray}
		\label{factorqcd}
		F^{\ttt}_{QCD}=\frac{g_{s}^{6}}{9}\delta_{AD}\delta_{BF}\delta_{EC}Tr(t^{A}t^{B}t^{C})\Big[\frac{1}{2}Tr\big(t^{D}t^{E}t^{F}\big)+\frac{1}{2}Tr\big(t^{D}t^{F}t^{E}\big)\Big]=\frac{g_{s}^{6}}{16\cdot9}d^{2} , \\
		\label{factorqed}
		F^{\ttt}_{QED}=n_{\ttt}\Big\{\frac{g_{s}^{4}e^{2}Q_{q}Q_{t}}{9}\delta_{AC}\delta_{BD}Tr(t^{A}t^{B})Tr(t^{C}t^{D})\Big\}=\frac{6g_{s}^{4}e^{2}}{9}Q_{t}Q_{q} .
	\end{eqnarray}
\end{subequations} 
$F^{\ttt}_{QCD}$ contains two different color structures and the result
depends on $d^2=d^{ABC}d_{ABC}=\frac{40}{3}$, which arises from
$Tr(t^{A}t^{B}t^{C})=\frac{1}{4}(i f^{ABC}+d^{ABC})$. 
$F^{\ttt}_{QED}$ instead depends on the charges of the incoming quarks ($Q_{q}$)
and of the top quark ($Q_{t}$), together with 
$n_{\ttt}=3$ corresponding to Figure~\ref{bornbox}.

\begin{figure}[!h]
\centering
\input{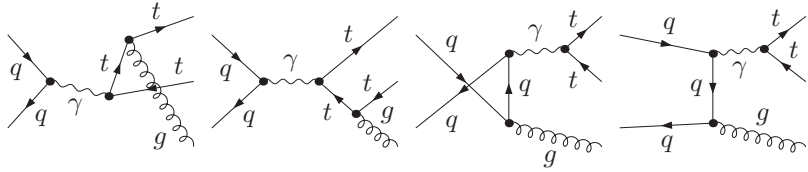}
\caption{Real gluon emission from photon exchange diagrams}
\label{realgqed}
\end{figure}

\begin{figure}[!h]
\centering
\input{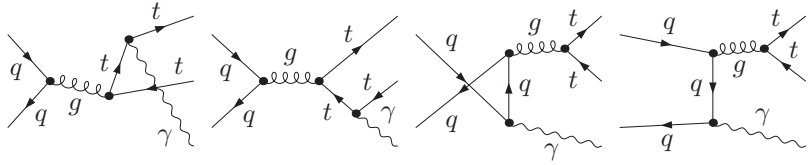}
\caption{Real photon emission from gluon exchange diagrams}
\label{realph}
\end{figure}

In a similar way, also the real-radiation processes
$q\bar{q}\rightarrow\ttt g$ and $q\bar{q}\rightarrow\ttt \gamma$ 
(Figures~\ref{realgqed} and \ref{realph})  can be
evaluated starting from the result obtained for $q\bar{q}\rightarrow\ttt g$
in the QCD case and  substituting successively each gluon by a photon, yielding
the ratios
\begin{equation}\label{rappqcdewttg}
\frac{\overline{|\mathcal{M}^{\ttt
      g}|}^{2}_{\mathcal{O}(\alphas^{2}\alpha),asym}}{\overline{|\mathcal{M}^{\ttt
      g}|}^{2}_{\mathcal{O}(\alphas^{3}),asym}}=\frac{\overline{2Re\big(\mathcal{M}^{\ttt
      g}_{\mathcal{O}(\alpha\sqrt{\alphas})}\mathcal{M}^{\ttt
      g\quad*}_{\mathcal{O}(\alphas\sqrt{\alphas})}\big)}_{asym}}{\overline{\big|\mathcal{M}^{\ttt
      g}_{\mathcal{O}(\alphas\sqrt{\alphas})}\big|}_{asym}^{2}}=\frac{F^{\ttt
    g}_{QED}(\alphas,\alpha,Q_{t},Q_{q})}{F^{\ttt g}_{QCD}(\alphas)} ,
\end{equation}
\begin{equation}\label{rappqcdewttgamma}
\frac{\overline{|\mathcal{M}^{\ttt
      \gamma}|}^{2}_{\mathcal{O}(\alphas^{2}\alpha),asym}}{\overline{|\mathcal{M}^{\ttt
      g}|}^{2}_{\mathcal{O}(\alphas^{3}),asym}}=\frac{\overline{\big|\mathcal{M}^{\ttt
      \gamma}_{\mathcal{O}(\alphas\sqrt{\alpha})}\big|}_{asym}^{2}}{\overline{\big|\mathcal{M}^{\ttt
      g}_{\mathcal{O}(\alphas\sqrt{\alphas})}\big|}_{asym}^{2}}=\frac{F^{\ttt
    \gamma}_{QED}(\alphas,\alpha,Q_{t},Q_{q})}{F^{\ttt g}_{QCD}(\alphas)} .
\end{equation} 
$F^{\ttt g}_{QCD}$, $F^{\ttt g}_{QED}$ and $F^{\ttt \gamma}_{QED}$ are related
to $F^{\ttt}_{QCD}$, $F^{\ttt}_{QED}$ 
in the following way,
\begin{subequations}
	\label{relations}
	\begin{eqnarray}
		\label{rel1}
		F^{\ttt g}_{QCD}= F^{\ttt }_{QCD},	\qquad	
                F^{\ttt g}_{QED} &=& \frac{2}{3}F^{\ttt }_{QED}, \qquad	
                F^{\ttt \gamma}_{QED}= \frac{1}{3}F^{\ttt }_{QED}, \\
		\label{rel2}
		F^{\ttt }_{QED}&=& F^{\ttt g }_{QED}+F^{\ttt \gamma }_{QED} .
	\end{eqnarray}
\end{subequations}
This guarantees the cancellation of the IR singularities stemming from the
virtual contributions.

The $\mathcal{O}(\alphas^{2}\alpha)$ antisymmetric term from $q\bar{q}\rightarrow\ttt g$ comes from
the interference of $q\bar{q}\rightarrow g \rightarrow \ttt g$
(Figure~\ref{realg}) and $q\bar{q}\rightarrow\gamma\rightarrow\ttt g$
(Figure~\ref{realgqed}). 
It can be obtained from the corresponding QCD result with the replacement 
of one gluon by a photon and the right couplings, as done in the case of
$q\bar{q} \rightarrow \ttt$. 
The only difference is the number of gluons to be replaced:
in the  $q\bar{q}\rightarrow \ttt g$ case they are only two instead of three
as for the virtual photon contributions.\\

The $\mathcal{O}(\alphas^{2}\alpha)$ antisymmetric term from $q\bar{q}\rightarrow\ttt \gamma$ 
comes from the $q\bar{q}\rightarrow g \rightarrow \ttt \gamma$ diagrams in Figure~\ref{realph}, 
and again it can be obtained from the corresponding QCD result for the
gluon-radiation process $q\bar{q}\rightarrow\ttt g$.  
Here we have a one-to-one relation between the QED and QCD diagrams.

Finally, we can relate the QED contribution
to the antisymmetric term $\tilde{N}_1$ in \eqref{powersew} to the
$\mathcal{O}(\alphas^{3})$ QCD term $N_1$ for a given quark species 
$q\bar{q}\rightarrow \ttt+X$ in the following way, 
\begin{equation}\label{rqed}
R_{QED}(Q_q) 
             = \frac{\alpha \tilde{N}_{1}^{QED}}{\alpha_s N_{1}}
             =\frac{F^{\ttt}_{QED}}{F^{\ttt}_{QCD}}
             = Q_{q}Q_{t}\frac{36}{5}\frac{\alpha}{\alphas} .
\end{equation}

Now we consider the weak contribution to $\tilde{N}_1$. 
It can be depicted by the same diagrams as for  
$q\bar{q}\rightarrow\ttt $ and $q\bar{q}\rightarrow\ttt g$ in the QED case, 
but with the photon now substituted by a $Z$ boson, involving massive box
diagrams.
The result cannot be expressed immediately in a simple factorized way. 
We performed the explicit calculation including also the contribution
from real gluon radiation with
numerical integration over the hard gluon part.

Basically also $Z$-boson radiation, $q\bar{q}\rightarrow\ttt Z$, can
contribute at the same order. As our calculation has shown, it yields 
only a tiny effect of $10^{-5}$ in $\afb$ and thus may be safely neglected.  
The same applies to $u\bar{d}\rightarrow\ttt W^{+}$ as well as to Higgs-boson
radiation.

Weak one-loop contributions to the $q\bar{q}g$ and  $t\bar{t}g$  vertices
induce also axialvector form factors, which however yield vanishing interference
terms with the Born amplitude for the antisymmetric cross section 
at $\mathcal{O}(\alphas^{2}\alpha)$ and are thus irrelevant.

\section{Numerical results}
The numerical analysis is based on the analytical evaluation of the required
symmetric and antisymmetric parts of the parton cross sections and
semi-numerical phase-space integration for the radiation processes 
with phase-space slicing, with support of
\textit{FeynArts}~\cite{Hahn:2000kx} and \textit{FormCalc}~\cite{Hahn:1998yk}.
This is done also for the QED subclass
starting from the $\ttt$, $\ttt g$ and $\ttt \gamma$ diagrams,
for comparison with the QED result obtained from \eqref{rqed},
showing perfect compatibility.

We choose MRST2004QED parton distributions \cite{Martin:2004dh} for NLO calculations and MRST2001LO
for LO \cite{Martin:2002dr}, 
using thereby $\alphas(\mu)$ of MRST2004QED also for the evaluation of the cross sections at LO
(a similar strategy was employed in~ \cite{Bernreuther:2010ny}).
The same value $\mu$ is used also for the factorization scale.
The numerical results are presented with three different choices for the
scale:  $\mu=m_{t}/2,m_{t},2m_{t}$. Other input parameters are taken from \cite{Nakamura:2010zzi}.

The results for the cross sections from the individual partonic channels and
their sum, yielding the denominator of $\afb$, are listed in~Table~\ref{cs}.
The various antisymmetric terms entering the numerator of either of the 
two variants $\att$ and $\alab$ are collected 
in Table~\ref{NN},   
and the corresponding contributions to the asymmetry in Table~\ref{NNperc}.
\begin{table}[!h]
\centering
\footnotesize
\begin{tabular}{r|rrr}
\hline
\hline
$\sigma$(pb)		&$\mu=m_{t}/2$	&$\mu=m_{t}$	&$\mu=2m_{t}$\\
\hline
\hline
$u\bar{u}$		& 6.245					& 4.454				& 3.355\\
\hline
$d\bar{d}$		& 1.112					& 0.777				& 0.575\\
\hline
$s\bar{s}$		& $1.37\times10^{-2}$	&$9.60\times10^{-3}$	& $0.706\times10^{-2}$\\
\hline
$c\bar{c}$		& $2.24\times10^{-3}$	&$1.69\times10^{-3}$	& $1.32\times10^{-3}$\\
\hline
$gg$			& 0.617					& 0.378				& 0.248\\
\hline
\hline
$p\bar{p}$		& 7.990					& 5.621				& 4.187\\
\hline
\hline
\end{tabular}
\caption{Integrated cross sections at $\mathcal{O}(\alphas^{2})$
          from the various partonic channels}\label{cs}
\end{table}

\normalsize

As already mentioned, the QED part was obtained in two different ways based on
the diagrammatic calculation and on the use of \eqref{rqed}; the weak part
results exclusively from the diagrammatic calculation.
The ratio of the total
$\mathcal{O}(\alphas^{2}\alpha)+\mathcal{O}(\alpha^{2})$ and
$\mathcal{O}(\alphas^{3})$ contributions to the numerator $N$
of the asymmetry \eqref{powersew} gives an illustration of the impact of the
electroweak relative to the QCD asymmetry.
The values obtained numerically for $\mu=(m_{t}/2,m_{t},2m_{t})$  
for the two definitions of $\afb$ are
\begin{eqnarray}
\label{rs}
R_{EW}^{\ttt} &=&
\frac{N^{\ttt}_{\mathcal{O}(\alphas^{2}\alpha)+\mathcal{O}(\alpha^{2})}}{N^{\ttt}_{\mathcal{O}(\alphas^{3})}}=(0.190,0.220,0.254), \nonumber\\
R_{EW}^{p\bar{p}} &=&
\frac{N^{p\bar{p}}_{\mathcal{O}(\alphas^{2}\alpha)+\mathcal{O}(\alpha^{2})}}{N^{p\bar{p}}_{\mathcal{O}(\alphas^{3})}}=(0.186,0.218,0.243), 
\end{eqnarray}
which are larger than the estimate of 0.09 given in \cite{Kuhn:1998kw}.
This shows that the electroweak contribution provides a non-negligible fraction
of the QCD-based antisymmetric cross section with the same overall sign, thus
enlarging the Standard Model prediction for the asymmetry
(the electroweak $\mathcal{O}(\alphas^{2}\alpha)$  contribution of
$u\bar{u}\rightarrow\ttt$  to the asymmetry is even bigger than the
$\mathcal{O}(\alphas^{3})$ contribution of $d\bar{d}\rightarrow\ttt$).
  
The final result for the two definitions of $\afb$  
can be summarized as follows,
\begin{equation}\label{afbvalues}
\att=(9.7,8.9,8.3)\% , \qquad \alab=(6.4,5.9,5.4)\% . 
\end{equation}

\begin{figure}[h]
\centering
\subfigure[$\att$\label{thextt}]{
\includegraphics[scale=0.55]{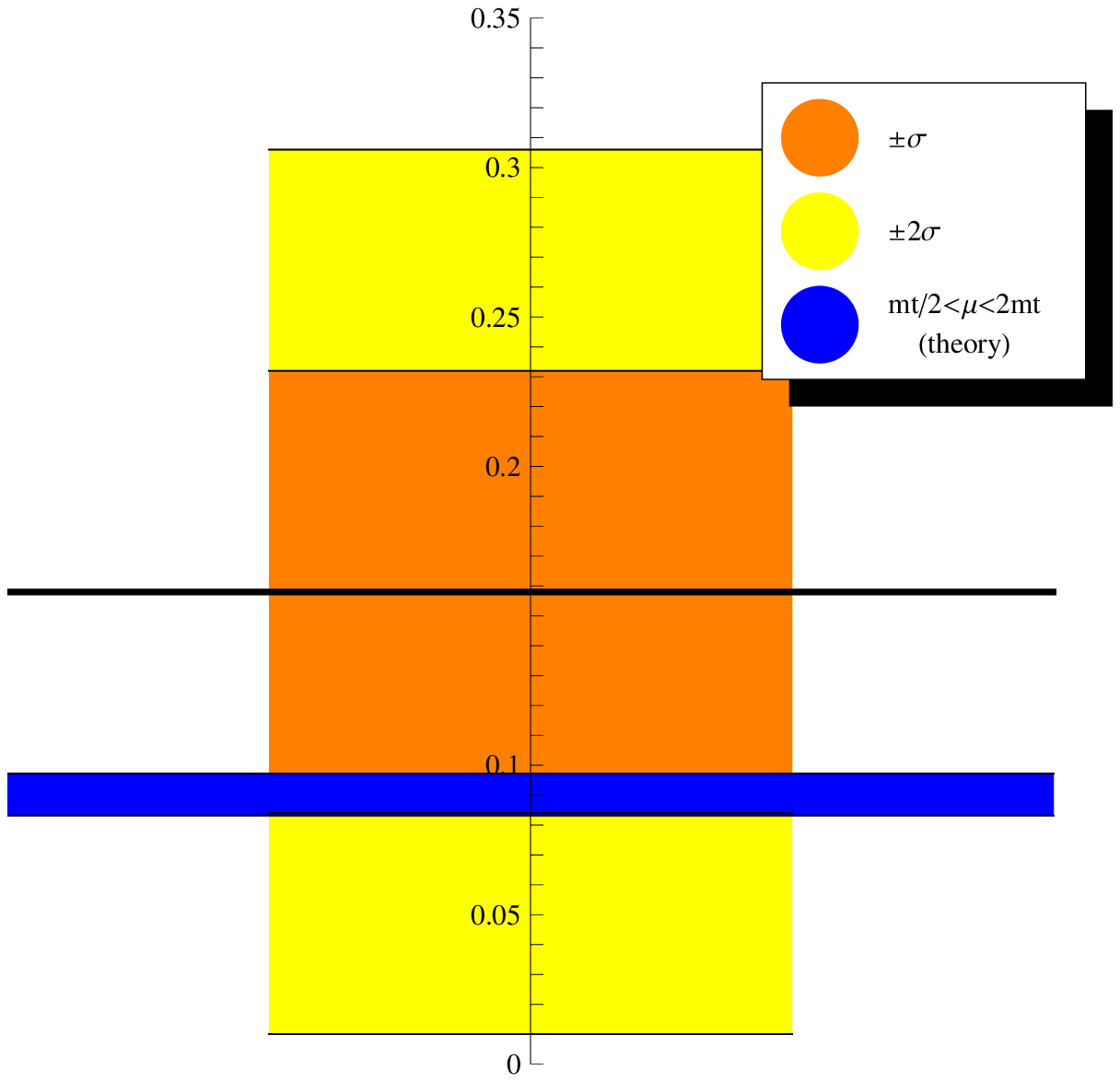}}
\subfigure[$\alab$\label{thexlab}]{
\includegraphics[scale=0.55]{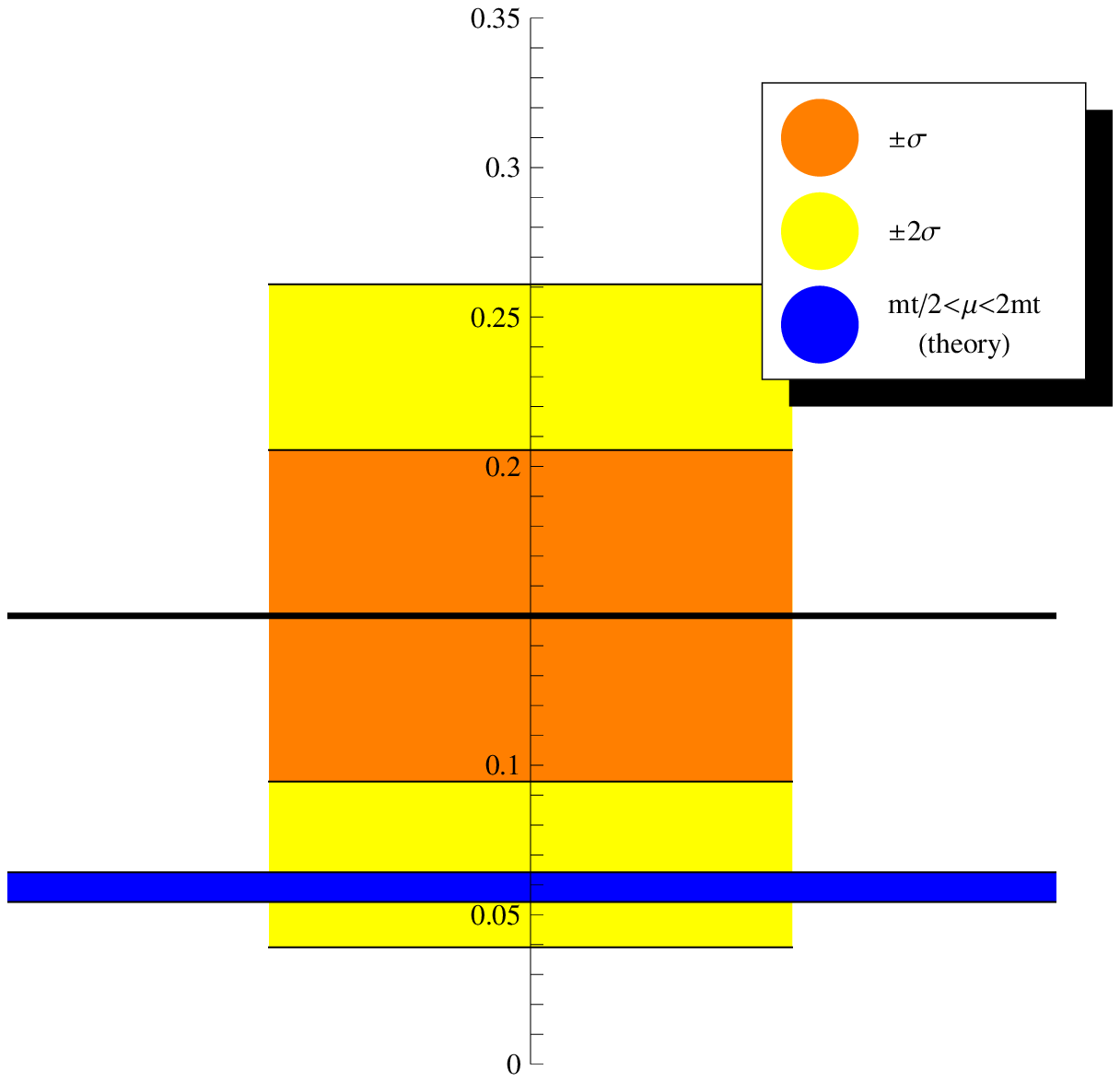}}
\caption{Theory(blue) and experimental data (black=central value, orange=$1\sigma$, yellow=$2\sigma$)}
\label{thex}
\end{figure}

Figure~\ref{thex} displays the theoretical prediction versus the experimental data. 
The prediction is almost inside the experimental $1\sigma$ range
for $\att$ and inside the $2\sigma$ range for $\alab$.
It is important to note that the band indicating the scale
variation of the prediction does not
account for all the theoretical uncertainties. 
For example, 
the $\mathcal{O}(\alphas^{4})$ term in $N$ is missing, and
we did not include the $\mathcal{O}(\alphas^{3})$ part in $D$. 
Including the NLO term for the cross section in D would decrease the
asymmetry by about 30\%, which indicates the size of the NLO terms.
In a conservative spirit one would consider this as an uncertainty
from the incomplete NLO calculation (see also the discussion
in~\cite{Kuhn:1998kw}).

We have performed our analysis also for applying two different types of cuts,
one to the $\ttt$ invariant mass and the other one to the rapidity:
$M_{\ttt}>450 \text{ GeV}$ and  $|\Delta y|>1$.
With those cuts, experimental data have also been presented 
in~\cite{Aaltonen:2011kc}. 
The cross section values for these cuts at LO are given in 
Table~\ref{csminvy}.
\begin{table}[!h]
\centering
\footnotesize
\begin{tabular}{r|rrr}
\hline
\hline
$\sigma$(pb)		&$\mu=m_{t}/2$	&$\mu=m_{t}$	&$\mu=2m_{t}$\\
\hline
\hline
$p\bar{p}(M_{\ttt}>450 \text{ GeV})$		& 3.113					& 2.148				& 1.573\\
\hline
$p\bar{p}(|\Delta y|>1)$		& 1.846					& 1.276				& 0.937\\
\hline
\hline
\end{tabular}
\caption{Cross sections with cuts at $\mathcal{O}(\alphas^{2})$}
\label{csminvy}
\end{table}
The various terms of the antisymmetric cross section contributing  
to $N$, as discussed above in the case without cuts, 
are now calculated for $\att$ 
for both cases $M_{\ttt}>450 \text{ GeV}$ and  $|\Delta y|>1$. 
The corresponding contributions to the asymmetry $\att$ are 
the entries of Table~\ref{NNpercminvy}.
The asymmetry with cuts is the total result,
\begin{equation}
\label{afbvaluesminvy}
\att(M_{\ttt}>450 \text{ GeV})=(13.9,12.8,11.9)\%, \qquad 
\att(|\Delta y|>1)=(20.7,19.1,17.5)\% .
\end{equation}
A comparison of Table~\ref{NNpercminvy} with Table~\ref{perctt} shows 
that the ratio of the QCD contribution to the  $u\bar{u}\rightarrow\ttt$ 
and $d\bar{d}\rightarrow\ttt$ subprocesses is larger with the $M_{\ttt}>450 \text{ GeV}$ cut,
 which leads to a slight increase of $R_{EW}^{\ttt}$:
\begin{equation}\label{rsminvy}
R_{EW}^{\ttt}(M_{\ttt}>450 \text{ GeV})=(0.200,0.232,0.266)\qquad R_{EW}^{\ttt}(|\Delta y|>1)=(0.191,0.216,0.246) . 
\end{equation}
It is, however, not enough to improve the situation.

\begin{figure}[h]
\centering
\subfigure[$\att(M_{\ttt}>450 \text{ GeV})$\label{thextt}]{
\includegraphics[scale=0.55]{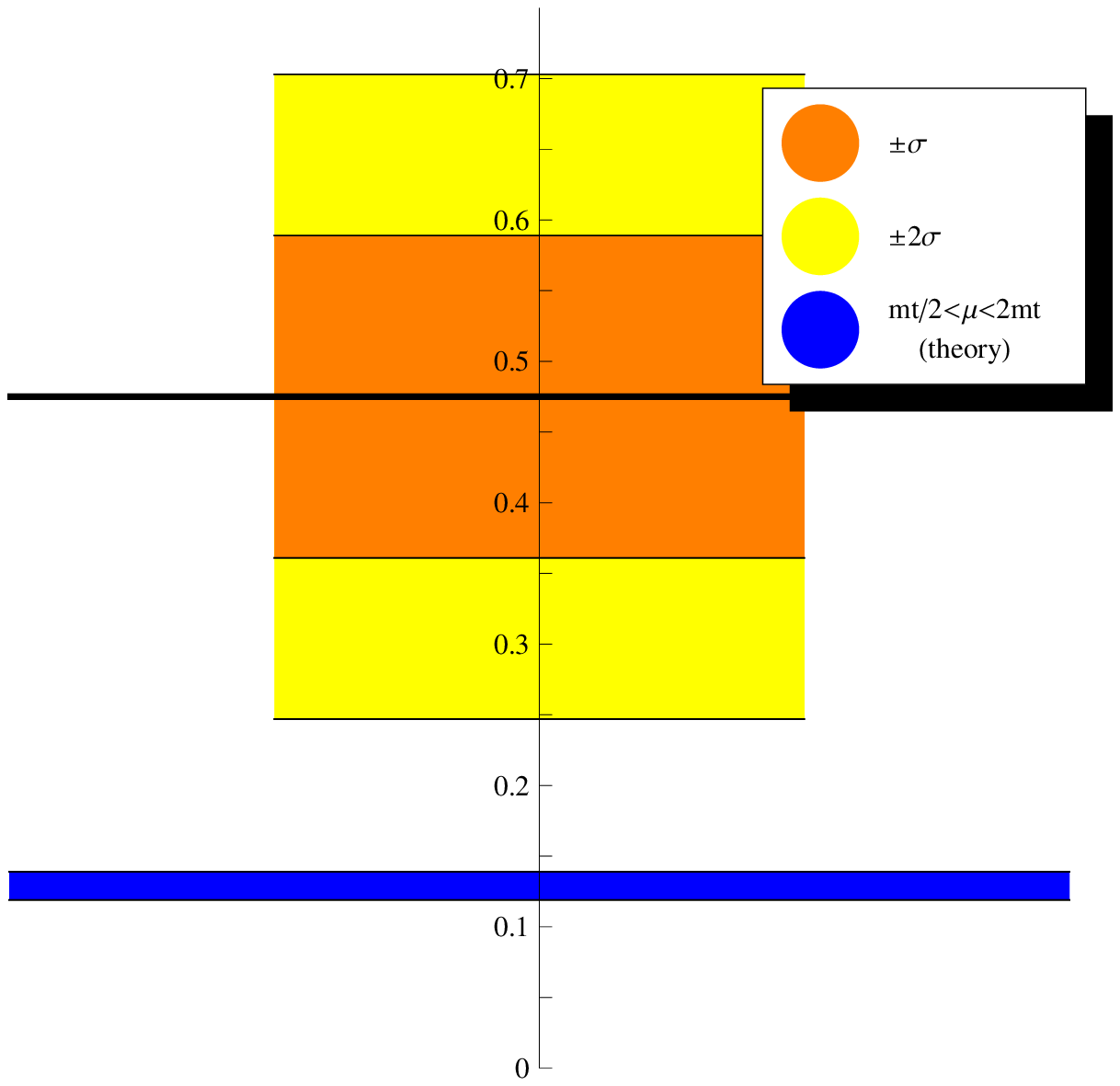}}
\subfigure[$\att(|\Delta y|>1)$\label{thexlab}]{
\includegraphics[scale=0.55]{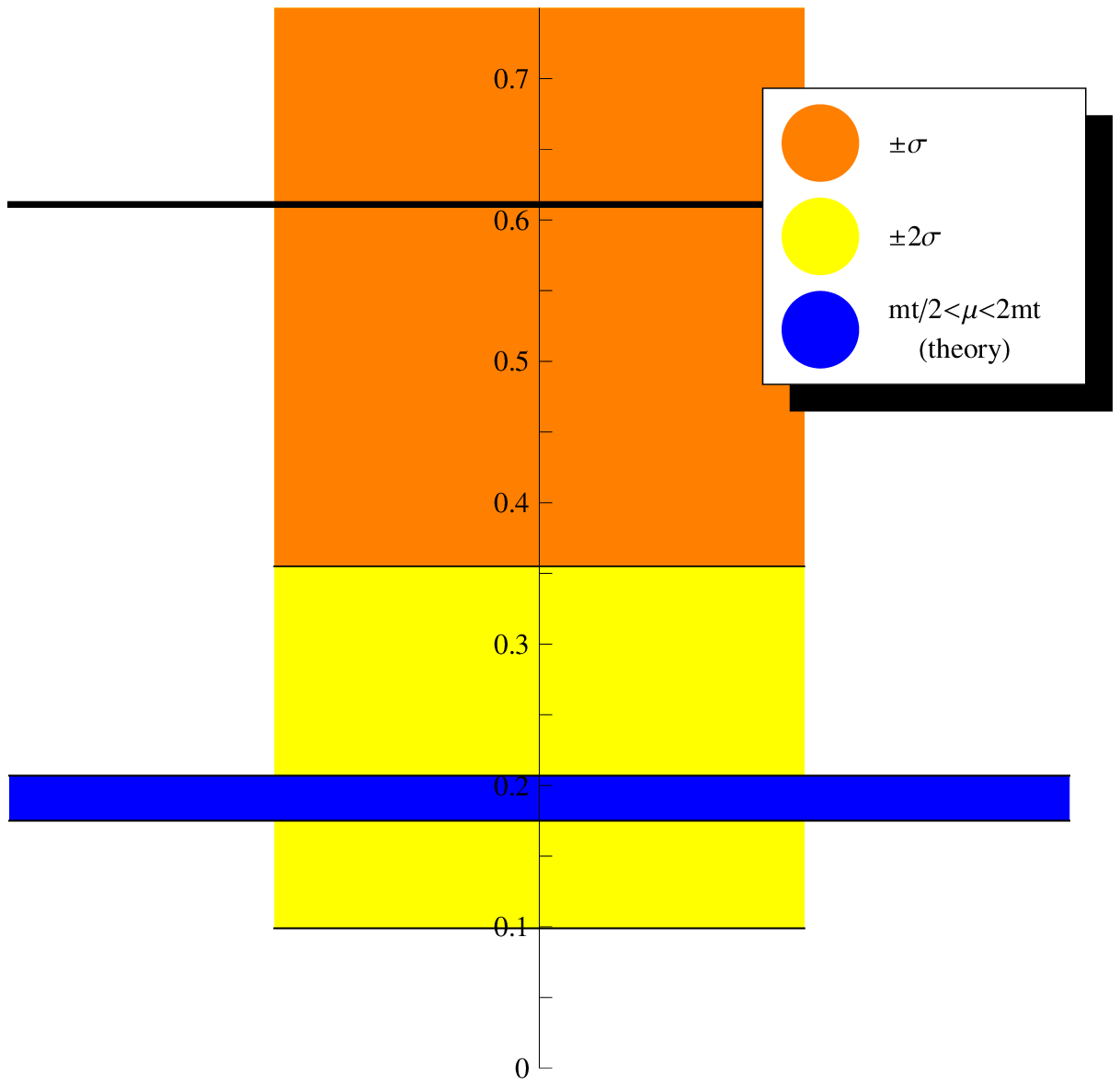}}
\caption{Theory(blue) and experimental data (black=central value, orange=$1\sigma$, yellow=$2\sigma$)}
\label{thexminvy}
\end{figure}

Figure~\ref{thexminvy} displays 
the theoretical prediction versus data for $\att$ with cuts. The Standard Model prediction 
is inside the $2\sigma$ range for the $|\Delta y|>1$ cut,  
but it is at the $3\sigma$ boundary for the invariant-mass cut $M_{\ttt}>450 \text{ GeV}$.

\section{Conclusions}
Our detailed analysis of the electroweak contributions  
to the forward-backward asymmetry in $\ttt$ production 
shows that they provide a non-negligible fraction
of the QCD-induced asymmetry with the same overall sign, thus
enlarging the Standard Model prediction for the asymmetry
at the Tevatron. For high invariant masses, a
$3\sigma$ deviation from the measured value still persists. 
The observed dependence of $\afb$ on the invariant mass of $\ttt$ 
could be an indication
for the presence of new physics below the TeV scale; it is, however,
difficult to interpret these deviations
as long as the NLO QCD calculation for the asymmetry is not available. \\

\begin{table}[h]
\centering
\subfigure[$\att$\label{Ntt}]{%
\footnotesize
\begin{tabular}{r|rrr}
\hline
\hline
$N$(pb)		&$\mu=m_{t}/2$	&$\mu=m_{t}$	&$\mu=2m_{t}$\\
\hline
\hline
$\mathcal{O}(\alphas^{3})\quad u\bar{u}$					& 0.560					& 0.354					& 0.234\\
\hline
$\mathcal{O}(\alphas^{3})\quad d\bar{d}$					& $9.25\times10^{-2}$	& $5.76\times10^{-2}$	& $3.76\times10^{-2}$\\
\hline
\hline
$\mathcal{O}(\alphas^{2}\alpha)_{QED}\quad u\bar{u}$		& 0.108				&0.0759					&0.0554\\
\hline
$\mathcal{O}(\alphas^{2}\alpha)_{QED}\quad d\bar{d}$		& $-8.9\times10^{-3}$		&$-6.2\times10^{-3}$		& $-4.5\times10^{-3}$\\
\hline
\hline
$\mathcal{O}(\alphas^{2}\alpha)_{weak}\quad u\bar{u}$		& $1.25\times10^{-2}$	& $0.89\times10^{-2}$	&  $0.66\times10^{-2}$\\
\hline
$\mathcal{O}(\alphas^{2}\alpha)_{weak}\quad d\bar{d}$		& $-3.6\times10^{-3}$		&$-2.5\times10^{-3}$		& $-1.8\times10^{-3}$\\
\hline
\hline
$\mathcal{O}(\alpha^{2})\quad u\bar{u}$					& $1.47\times10^{-2}$			& $1.30\times10^{-2}$&  $1.17\times10^{-2}$\\
\hline
$\mathcal{O}(\alpha^{2})\quad d\bar{d}$					& $1.8\times10^{-3}$		&$1.6\times10^{-3}$		& $1.4\times10^{-3}$\\
\hline
\hline
\end{tabular}
}

\subfigure[$\alab$\label{Nlab}]{%
\footnotesize
\begin{tabular}{r|rrr}
\hline
\hline
$N$(pb)		&$\mu=m_{t}/2$	&$\mu=m_{t}$	&$\mu=2m_{t}$\\
\hline
\hline
$\mathcal{O}(\alphas^{3})\quad u\bar{u}$					&0.373					& 0.236					& 0.155\\
\hline
$\mathcal{O}(\alphas^{3})\quad d\bar{d}$					& $5.97\times10^{-2}$	& $3.72\times10^{-2}$	& $2.42\times10^{-2}$\\
\hline
\hline
$\mathcal{O}(\alphas^{2}\alpha)_{QED}\quad u\bar{u}$		& $7.15\times10^{-2}$	&$5.06\times10^{-2}$		&$3.67\times10^{-2}$\\
\hline
$\mathcal{O}(\alphas^{2}\alpha)_{QED}\quad d\bar{d}$		& $-5.7\times10^{-3}$		&$-4.0\times10^{-3}$		& $-2.9\times10^{-3}$\\
\hline
\hline
$\mathcal{O}(\alphas^{2}\alpha)_{weak}\quad u\bar{u}$		& $8.2\times10^{-3}$		& $5.8\times10^{-3}$		&  $4.2\times10^{-3}$\\
\hline
$\mathcal{O}(\alphas^{2}\alpha)_{weak}\quad d\bar{d}$		& $-2.3\times10^{-3}$		&$-1.6\times10^{-3}$		& $-1.1\times10^{-3}$\\
\hline
\hline
$\mathcal{O}(\alpha^{2})\quad u\bar{u}$					& $9.1\times10^{-3}$		& $8.0\times10^{-3}$		&  $7.1\times10^{-3}$\\
\hline
$\mathcal{O}(\alpha^{2})\quad d\bar{d}$					& $1.1\times10^{-3}$		&$1.0\times10^{-3}$		& $0.9\times10^{-3}$\\
\hline
\hline
\end{tabular}
}
\caption{The various contributions to the antisymmetric cross section $N$ of $\att$ and $\alab$\label{NN}}
\end{table}

\clearpage
\begin{table}[h]
\centering
\subfigure[$\att$\label{perctt}]{%
\footnotesize
\begin{tabular}{r|rrr}
\hline
\hline
$\att$		&$\mu=m_{t}/2$	&$\mu=m_{t}$	&$\mu=2m_{t}$\\
\hline
\hline
$\mathcal{O}(\alphas^{3})\quad u\bar{u}$					& 7.01\%		& 6.29\%		& 5.71\%\\
\hline
$\mathcal{O}(\alphas^{3})\quad d\bar{d}$					& 1.16\%		& 1.03\%		&0.92\%\\
\hline
\hline
$\mathcal{O}(\alphas^{2}\alpha)_{QED}\quad u\bar{u}$		& 1.35\%		&1.35\%			&1.35\%	\\
\hline
$\mathcal{O}(\alphas^{2}\alpha)_{QED}\quad d\bar{d}$		& -0.11\%		&-0.11\%		& -0.11\%\\
\hline
\hline
$\mathcal{O}(\alphas^{2}\alpha)_{weak}\quad u\bar{u}$		& 0.16\%		&0.16\%			&  0.16\%	\\
\hline
$\mathcal{O}(\alphas^{2}\alpha)_{weak}\quad d\bar{d}$		& -0.04\%		&-0.04\%		& -0.04\%	\\
\hline
\hline
$\mathcal{O}(\alpha^{2})\quad u\bar{u}$					& 0.18\%		& 0.23\%		& 0.28\%\\
\hline
$\mathcal{O}(\alpha^{2})\quad d\bar{d}$					& 0.02\%		&0.03\%			& 0.03\%\\
\hline
\hline
$\text{tot}\quad p\bar{p}$ 											& 9.72\%		&8.93\%			&8.31\%\\
\hline
\hline
\end{tabular}
}
\subfigure[$\alab$\label{perclab}]{%
\footnotesize
\begin{tabular}{r|rrr}
\hline
\hline
$\alab$		&$\mu=m_{t}/2$	&$\mu=m_{t}$	&$\mu=2m_{t}$\\
\hline
\hline
$\mathcal{O}(\alphas^{3})\quad u\bar{u}$					&4.66\%			& 4.19\%		& 3.78\%\\
\hline
$\mathcal{O}(\alphas^{3})\quad d\bar{d}$					& 0.75\%		& 0.66\%		& 0.59\%\\
\hline
\hline
$\mathcal{O}(\alphas^{2}\alpha)_{QED}\quad u\bar{u}$		& 0.90\%		&0.90\%			&0.90\%\\
\hline
$\mathcal{O}(\alphas^{2}\alpha)_{QED}\quad d\bar{d}$		& -0.07\%		&-0.07\%		& -0.07\%\\
\hline
\hline
$\mathcal{O}(\alphas^{2}\alpha)_{weak}\quad u\bar{u}$		& 0.10\%		& 0.10\%		& 0.10\%\\
\hline
$\mathcal{O}(\alphas^{2}\alpha)_{weak}\quad d\bar{d}$		& -0.03\%		&-0.03\%		& -0.03\%\\
\hline
\hline
$\mathcal{O}(\alpha^{2})\quad u\bar{u}$					& 0.11\%		&0.14\%			&  0.17\%\\
\hline
$\mathcal{O}(\alpha^{2})\quad d\bar{d}$					& 0.01\%		&0.02\%			& 0.02\%\\
\hline
\hline
$\text{tot}\quad p\bar{p}$ 											& 6.42\%		&5.92\%			&5.43\%\\
\hline
\hline
\end{tabular}
}
\caption{Individual and total contributions to $\att$ and $\alab$\label{NNperc}}
\end{table}

\clearpage
\begin{table}[h]
\centering
\subfigure[$\att(M_{\ttt}>450 \text{ GeV})$\label{percminv}]{%
\footnotesize
\begin{tabular}{r|rrr}
\hline
\hline
$\att$		&$\mu=m_{t}/2$	&$\mu=m_{t}$	&$\mu=2m_{t}$\\
\hline
\hline
$\mathcal{O}(\alphas^{3})\quad u\bar{u}$					& 10.13\%		& 9.10\%		& 8.27\%\\
\hline
$\mathcal{O}(\alphas^{3})\quad d\bar{d}$					& 1.44\%		& 1.27\%		&1.14\%\\
\hline
\hline
$\mathcal{O}(\alphas^{2}\alpha)_{QED}\quad u\bar{u}$		& 1.94\%		&1.95\%			&1.96\%	\\
\hline
$\mathcal{O}(\alphas^{2}\alpha)_{QED}\quad d\bar{d}$		& -0.14\%		&-0.14\%		& -0.14\%\\
\hline
\hline
$\mathcal{O}(\alphas^{2}\alpha)_{weak}\quad u\bar{u}$		& 0.28\%		&0.28\%			&  0.28\%	\\
\hline
$\mathcal{O}(\alphas^{2}\alpha)_{weak}\quad d\bar{d}$		& -0.05\%		&-0.05\%		& -0.05\%	\\
\hline
\hline
$\mathcal{O}(\alpha^{2})\quad u\bar{u}$					& 0.26\%		& 0.33\%		& 0.41\%\\
\hline
$\mathcal{O}(\alpha^{2})\quad d\bar{d}$					& 0.03\%		&0.03\%			& 0.04\%\\
\hline
\hline
$\text{tot}\quad p\bar{p}$ 											& 13.90\%		&12.77\%			&11.91\%\\
\hline
\hline
\end{tabular}
}
\subfigure[$\att(|\Delta y|>1)$\label{percy}]{%
\footnotesize
\begin{tabular}{r|rrr}
\hline
\hline
$\att$		&$\mu=m_{t}/2$	&$\mu=m_{t}$	&$\mu=2m_{t}$\\
\hline
\hline
$\mathcal{O}(\alphas^{3})\quad u\bar{u}$					&15.11\%			& 13.72\%		& 12.41\%\\
\hline
$\mathcal{O}(\alphas^{3})\quad d\bar{d}$					&2.28\%		& 2.02\%		& 1.84\%\\
\hline
\hline
$\mathcal{O}(\alphas^{2}\alpha)_{QED}\quad u\bar{u}$		& 2.90\%		&2.94\%			&2.94\%\\
\hline
$\mathcal{O}(\alphas^{2}\alpha)_{QED}\quad d\bar{d}$		& -0.22\%		&-0.22\%		& -0.22\%\\
\hline
\hline
$\mathcal{O}(\alphas^{2}\alpha)_{weak}\quad u\bar{u}$		& 0.25\%		& 0.25\%		& 0.26\%\\
\hline
$\mathcal{O}(\alphas^{2}\alpha)_{weak}\quad d\bar{d}$		& -0.09\%		&-0.09\%		& -0.08\%\\
\hline
\hline
$\mathcal{O}(\alpha^{2})\quad u\bar{u}$					& 0.35\%		&0.45\%			&  0.55\%\\
\hline
$\mathcal{O}(\alpha^{2})\quad d\bar{d}$					& 0.04\%		&0.05\%			& 0.06\%\\
\hline
\hline
$\text{tot}\quad p\bar{p}$ 											& 20.70\%		&19.12\%			&17.75\%\\
\hline
\hline
\end{tabular}
}
\caption{Individual and total contributions to $\att(M_{\ttt}>450 \text{ GeV})$ and $\att(|\Delta y|>1)$\label{NNpercminvy}}
\end{table}
\pagebreak
\clearpage
\bibliography{biblio}
\bibliographystyle{JHEP}

\end{document}